\documentclass{ws-ijmpd}

\begin{document}

\markboth{D. Ma and P. He} {Distribution Function of Dark Matter}

%
\catchline{}{}{}{}{}
%

\title{Distribution Function of Dark Matter with Constant Anisotropy}

\author{Ding Ma\footnote{Email: mading@itp.ac.cn}}

\author{Ping He\footnote{Email: hep@itp.ac.cn}}

\address{Institute of Theoretical Physics, \\
Chinese Academy of Sciences, P. O. Box 2735,\\
 Beijing 100080,
China}

\maketitle

\begin{history}
\received{Day Month Year}
\revised{Day Month Year}
\comby{Managing Editor}
\end{history}

\begin{abstract}
N-body simulations of dark matter halos show that the density is
cusped near the center of the halo. The density profile behaves as
$r^{-\gamma}$ in the inner parts, where $\gamma \simeq 1$ for the
NFW model and $\gamma \simeq 1.5$ for the Moore's model, but in the
outer parts, both models agree with each other in the asymptotic
behavior of the density profile. The simulations also show the
information about anisotropy parameter $\beta(r)$ of velocity
distribution. $\beta\approx 0$ in the inner part and $\beta\approx
0.5$ (radially anisotropic) in the outer part of the halo. We
provide some distribution functions $F(E,L)$ with the constant
anisotropy parameter $\beta$ for the two spherical models of dark
matter halos: a new generalized NFW model and a generalized Moore
model. There are two parameters $\alpha$ and $\epsilon$ for those
two generalized models to determine the asymptotic behavior of the
density profile. In this paper, we concentrate on the situation of
$\beta(r)=1/2$ from the viewpoint of the simulation.

\end{abstract}


\section{Introduction}\label{S1}

The dark matter halo can be considered as the collisionless
self-gravitational system which is described by the Vlasov equation
and the Poisson equation. Apart from observations, there are mainly
two approaches to the study of dark matter halos: (1) numerical
simulations and (2) analytic or semi-analytic method.

Nowadays, N-body simulations become more and more important in the
study of dark matter halos. Those simulations provide density
profile and other properties of dark matter halos such as anisotropy
of velocity distribution. A 'universal' density profile of dark
matter halo was proposed by Navarro, Frenk, $\&$ White (hereafter
NFW)\cite{NFW95,NFW96}. The NFW profile is shown as $\rho\propto
1/((r/r_s)(1+r/r_s)^{2})$, where $r_s$ is a characteristic radius.
We can find directly from the NFW profile that $\rho\propto r^{-1}$
near the center and $\rho\propto r^{-3}$ in the outer parts. Some
simulations' results exhibit different inner logarithmic slopes from
NFW's but deviate not so much from the latter's in the outer parts
of the halo. In the inner parts of the halo, Moore's
profile\cite{M99} behaves as $\rho\propto r^{-1.5}$. Jing and Suto's
profile\cite{J2000} behaves as $\rho\propto r^{-\gamma}$ in the
inner part, where the density logarithmic slopes $\gamma\simeq 1.1
\sim 1.5$ with different merger histories and different total mass
of the halos. Some other simulations concentrate on the outer slope
of the dark matter density. For example, Avila-Reese et al. found
that outer slopes $\gamma$ of some halos are larger than NFW's outer
slopes($\gamma=3$)\cite{AR99}. What's more, Hansen and
Moore\cite{HM06,HS06} found that the density logarithmic slopes
$\gamma$ is correlated with the velocity anisotropy which is
parameterized as an anisotropy parameter $\beta$ and they provided
the formula $\beta\approx1-1.15(1-\gamma/6)$. So $\beta\approx 0$ in
the inner part as $\gamma\approx 1$ and $\beta\approx 0.5$ in the
outer part as $\gamma\approx 3$.

On the other hand, many authors try to construct the dynamical
models of the stellar system and the dark matter halo analytically
or semi-analytically. It is important to construct the dynamical
models which have physical meaning, as for instance, these analytic
or semi-analytic dynamical models can be used to generate the
initial conditions of N-body simulations\cite{BV07}. A stellar
system or a dark matter halo is described by the distribution
functions(DFs) \textbf{F(x,v)}. Eddington (1916) showed how to
determine the DF of a spherical symmetric stellar system with the
isotropic velocity distribution\cite{Ed16}, but it is difficult to
calculate the distribution function in the anisotropic cases. A
pioneer work on the anisotropic cases is called King-Michie
model\cite{M63,King66} that comes from an approximate steady state
solution of the Fokker-Planck equation. In the past few years, there
has been great progress in the anisotropic cases. Dejonghe took a
large step on finding anisotropic distribution function with using
the augmented density $\rho(\psi,r)$\cite{De86,De87}. Some authors
constructed the anisotropic models with constant anisotropy
parameter\cite{EA06}. For another kind of models called the
Osipkov-Merritt models\cite{Os79,Os79tr,Me85}, the velocity
dispersion tensor becomes isotropic near the center and becomes
completely radial anisotropic in the outer part of the halo.
Cuddeford\cite{Cu91} constructed the modified Osipkov-Merritt model
which allows the velocity dispersion tensor to be an arbitrary
anisotropy in the inner part of the halo but still completely radial
anisotropy in the outer part and the composite Osipkov-Merritt model
was constructed by Ciotti $\&$ Pellegrini\cite{CP92}. Furthermore,
Baes $\&$ Hese\cite{BV07} constructed a dynamical model with seven
parameters very recently to get a flexible anisotropy profile. Some
authors built a model\cite{TLS06} to erase the cusp in the center of
the halo.

To construct the dynamical models(or DFs) with flexible anisotropy
behavior, some special potential-density pairs have been considered.
For example, the Plummer model\cite{De87,Pl11}, anisotropic Veltmann
model\cite{Ve79}, the Hernquist model\cite{He90,BD02} and the
$\gamma-$model\cite{Deh93,Tre94,BDB05,BHD07}. Some authors study
this matter in other ways. Some assumed simple DFs at first and then
solved the the potential and the density profiles\cite{Too82,AE05}.
Widrow\cite{Wi00} provided the Osipkov-Merritt DFs and DEDs
(differential energy distributions) for the NFW profile by the
semi-analytic method.

Recently, Evans and An\cite{EA06} have provided the DFs with
constant anisotropy of the dark matter for two types of the density
profiles. One is the generalized NFW profiles, $\rho\propto
1/(r(a+r)^{b-1})$, and the other is the Gamma model, $\rho\propto
1/(r^\gamma(a+r)^{4-\gamma})$. Both in these two models, asymptotic
behaviors of density profiles are controlled by only one parameter.
In this paper, we provide the DFs with constant anisotropy of the
dark matter for more generalized density profiles in a
semi-analytical way. Our generalized density profiles, with two
parameters $\alpha$ and $\epsilon$, can cover many realistic
profiles which come from simulations.

In Section 2 we review the basic knowledge needed for this paper and
the main ideas of DF with constant anisotropy. Section 3 provides
DFs of two models for dark matter halos: a new generalized NFW model
and a generalized Moore model. We make the discussion and conclusion
in Section 4.

\section{Basic Properties}
\subsection{General formulae}

For the spherically symmetric stellar system with the isotropic
velocity field, a mass distribution function $F(E)$ describes this
system very well. The mass density can be obtained from the
distribution function\cite{De86},
\begin{equation}
 \rho(\psi)=4\pi \int_0^{\psi}F(E)\sqrt{2(\psi-E)}dE\ ,
\label{rhoiso}
\end{equation}
where the binding energy $E$ is defined as
\begin{equation}
E=\psi(r) - \tfrac{1}{2}\,v_r^2 - \tfrac{1}{2}\,v_T^2\ ,
\end{equation}
and $\psi(r)$ is the relative gravitational potential which can be
obtained from the Poisson equation:
\begin{equation}
  \frac{1}{r^2}\,\frac{d}{dr}
  \left(r^2\,\frac{d\psi}{dr}\right)
  =-4\pi G\rho(r)\ .
  \label{eq:poisson}
\end{equation}
$v_r$ is the radial velocity and $v_T$ is the tangential velocity:
\begin{equation}
  v_T=\sqrt{v_\theta^2+v_\varphi^2}\ .
\end{equation}
Eddington\cite{Ed16} provided the inversion formula of the
Eq.~(\ref{rhoiso}):
\begin{equation}
F(E)=\frac{1}{2\pi^2}D_E\int_0^E\frac{d\rho(\psi)}{d\psi}
\frac{d\psi}{\sqrt{2(E-\psi)}}\ , \label{Fiso}
\end{equation}
where $D_E$ denotes the differentiation with respect to $E$.

The anisotropy parameter\cite{Bi80} mentioned in the Introduction is
defined as:
\begin{equation}
\beta=1-\frac{\sigma_T^2}{2 \sigma_r^2}\ ,
\end{equation}
where $\sigma_T^2$ and $\sigma_r^2$ are the tangential and radial
velocity dispersion. If $\beta<0$, $\sigma_T^2>2\sigma_r^2$
(tangentially anisotropic). If $0<\beta\leq1$,
$\sigma_T^2<2\sigma_r^2$ (radially anisotropic). Else If $\beta=0$,
$\sigma_T^2=2\sigma_r^2$ (isotropic case). Specially,
$\beta\rightarrow-\infty$ means that every particle is in a circular
orbit and $\beta=1$ indicate that every particle is in a radial
orbit.

The anisotropic case is different from the isotropic one as we
should consider the modulus of the angular momentum
vector\cite{De86} $L=rv_T$ and therefore now the distribution
function of dark matter depends on two variables $E$ and $L$. Then,
the mass density can be obtained from $F(E,L)$ by the double
integration\cite{De86}
\begin{equation}
\rho(\psi,r)=2\pi \int_0^{\psi}dE\int_0^{2(\psi-E)}
             \frac{F(E,L)}{\sqrt{2(\psi-E)-v_T^2}}dv_T^2\ .
\label{eq:rhof}
\end{equation}
However, the inversion formula of the above equation is much more
difficult to obtain than that in the isotropic case. Many works have
been done for this problem just as mentioned in the Introduction.

\subsection{Distribution function with constant anisotropy}
There is a simple and widely used ansatz for $F(E,L)$ as
below\cite{De86,EA06,Cu91,AE06,WE99}:
\begin{equation}
F(E,L)=L^{-2\beta}f(E)\ . \label{eq:ansatz}
\end{equation}
The simulations indicate that the anisotropy parameter $\beta(r)$
varies radially , but for the DF above, $\beta(r)$ is constant.
Although this simple ansatz is very attractive in the calculation of
inversion formula of Eq.~(\ref{eq:rhof}), the assumption of constant
anisotropy parameter is not so ideal. Anyway, we can still use this
valuable assumption which pave the way for the further work.

 From Eq.~(\ref{eq:rhof}) and Eq.~(\ref{eq:ansatz}) we can
get $\rho$ from $F(E,L)$,
\begin{equation}
\rho(\psi,r)=r^{-2\beta}\,
\frac{(2\pi)^{3/2}\Gamma(1-\beta)}{2^\beta\Gamma(3/2-\beta)}
\int_0^\psi\!(\psi-E)^{1/2-\beta}f(E)\,dE\ . \label{eq:den}
\end{equation}
The function $f(E)$ can be given from the
 inversion formula\cite{EA06,Cu91,AE06} of the above equation.
\begin{equation}
 f(E)=
\frac{2^\beta(2\pi)^{-3/2}}{\Gamma(1-\lambda)\Gamma(1-\beta)}\
\frac{d}{dE}\!\int_0^E\frac{d\psi}{(E-\psi)^\lambda}
\frac{d^nh}{d\psi^n}\ , \label{eq:disint}
\end{equation}
where $h(\psi)=r^{2\beta}\rho$ is expressed as a function of $\psi$,
$n=\lfloor(3/2-\beta)\rfloor$ and $\lambda=3/2-\beta-n$ is the
integer floor and the fractional part of $3/2-\beta$. The DED now
can be expressed as \cite{EA06,Cu91}
\begin{equation}
\frac{dM}{dE}=f(E)\,
\frac{(2\pi)^{5/2}\Gamma(1-\beta)}{2^{\beta-1}\Gamma(3/2-\beta)}
\int_0^{r_E}(\psi-E)^{1/2-\beta}r^{2(1-\beta)}dr\ , \label{eq:ded}
\end{equation}
where $r_E$ is defined through $\psi(r_E)=E$. It is clear that $r_E$
is the largest radius reachable by a particle with binding energy E.

The expressions of DFs and DEDs are relatively simple if $\beta$ is
a half integer (i.e., $\beta=3/2, 1/2,-1/2...$). Specifically, in
the case of $\beta=1/2$, which is more suitable from the viewpoint
of the simulation, the expressions of DF and DED further reduce
to\cite{EA06}
\begin{gather}
\begin{split}&
F(E,L)=\frac{g(r_E)}{2\pi^2L}\,,\qquad \frac{dM}{dE}=2\pi
r_E^2g(r_E), \label{eq:def}
\\&
g(r_E)=\left.\frac{\rho+r(d\rho/dr)}{(d\psi/dr)}\right|_{r=r_E}.
\end{split}
\end{gather}
In this paper, we select $\beta=1/2$ for our models just as
Ref.~\refcite{EA06} by the same token.

\section{Dark Matter Halo Models}

\subsection{New generalized NFW profiles}

Let us consider a family of density profiles with parameters
$\alpha$ and $\epsilon$
\begin{equation}
\rho=C\frac{1}{(r/r_s)^\alpha(1+r/r_s)^{3+\epsilon-\alpha}}
 \label{eq:rho}\ .
\end{equation}

We set the characteristic radius $r_s=1$, the total mass $M_{tot}=1$
and the gravitational constant $G=1$ here, and then the density
profiles reduce to
\begin{equation}
\rho=C\frac{1}{r^\alpha(1+r)^{3+\epsilon-\alpha}}
 \label{eq:rho}\ ,
\end{equation}

\begin{equation}
C=\frac{\Gamma(3-\alpha+\epsilon)}{4\pi\Gamma(3-\alpha)\Gamma(\epsilon)}\
.
\end{equation}
With two parameters $\alpha<3$ and $\epsilon>0$ we can freely select
the asymptotic behavior of the density profile both in the inner and
outer part of the halo. For the cases of $0<\epsilon\ll1$, the
profile closes to the NFW profiles when $\alpha\rightarrow1$ and is
similar to the Moore or Jing \& Suto's profile if
$\alpha=1.1\sim1.5$. If $\epsilon=1$, the profile just reduces to
the $\gamma$ model.

We can calculate the relative potential $\psi$ from the Poisson
equation.
\begin{gather}
\begin{split}&
 \psi(r)=C\left[c_0- 4\pi
 r^{-\alpha}\Gamma(-\alpha)\left(A_1(r)-2A_2(r)+A_3(r)\right)\right]
\\&
\\&
 A_1(r)=\ _2\tilde{F}_1[1-\alpha+\epsilon,\ -\alpha,\ 2-\alpha,\ -r]
\\&
 A_2(r)=\ _2\tilde{F}_1[2-\alpha+\epsilon,\ -\alpha,\ 2-\alpha,\ -r]
\\&
 A_3(r)=\ _2\tilde{F}_1[3-\alpha+\epsilon,\ -\alpha,\ 2-\alpha,\ -r]
 \label{eq:psi}
\end{split}
\end{gather}
where $c_0$ is determined by the condition
$\left.\psi(r)\right|_{r\rightarrow\infty}=0$. Note that
$\Gamma(-\alpha)$ is singular when $\alpha=1, 2$. We find that
$\psi$ cannot satisfy the condition
$\left.\psi(r)\right|_{r\rightarrow\infty}=0$ when
$\alpha\rightarrow 2$, but there is no problem for
$\alpha\rightarrow 1$. In this paper, we focus on the situation of
$\alpha=1 \sim 1.5$ from the viewpoint of the N-body simulation.

\begin{figure*}[h]
\includegraphics[width=0.482\hsize]{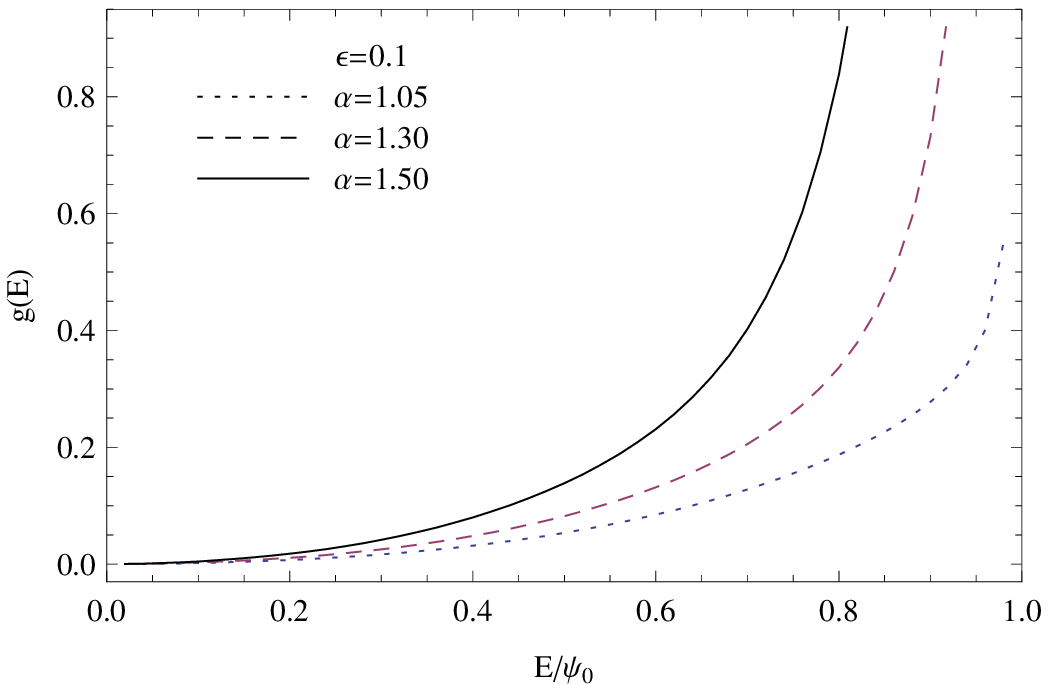}
\includegraphics[width=0.482\hsize]{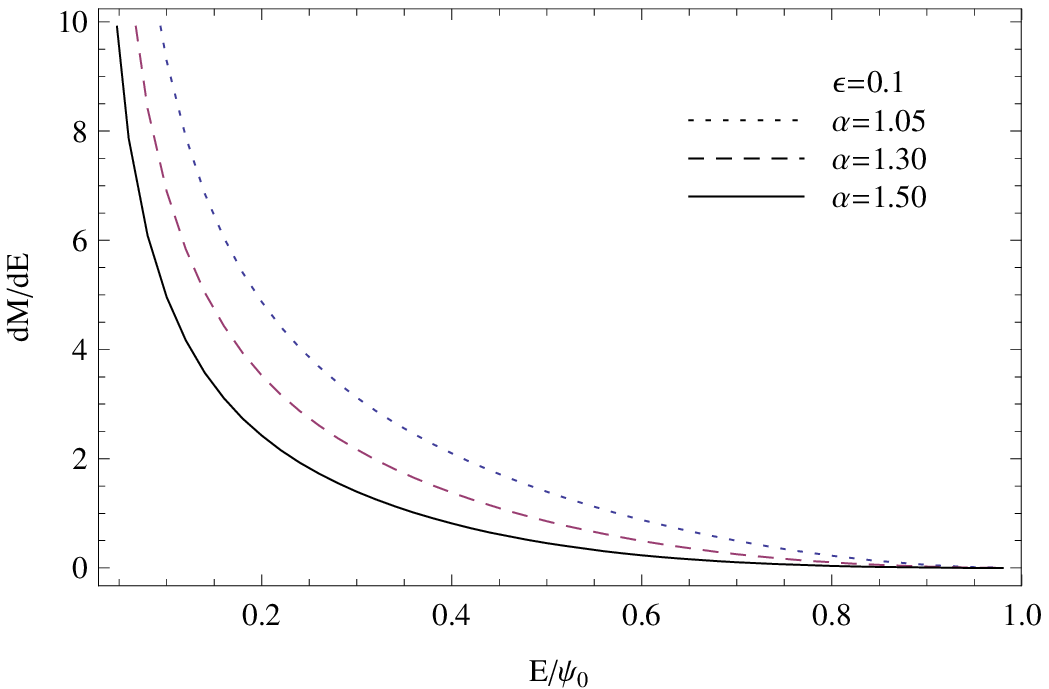}
\vspace*{8pt}
 \caption{\label{fig:Fig1}\footnotesize The energy part
of the distribution function (left panel) and the differential
energy distribution (right panel) of the New Generalized NFW model
with the constant anisotropy parameter $\beta=1/2$ and
$\epsilon=0.1$: dotted lines ($\alpha=1.05$, closes to the NFW
model), dashed lines ($\alpha=1.30$), solid lines ($\alpha=1.50$,
where the asymptotic behavior of density profile is similar to the
Moore model). \label{f1}}
\end{figure*}

\begin{figure*}[h]
\includegraphics[width=0.48\hsize]{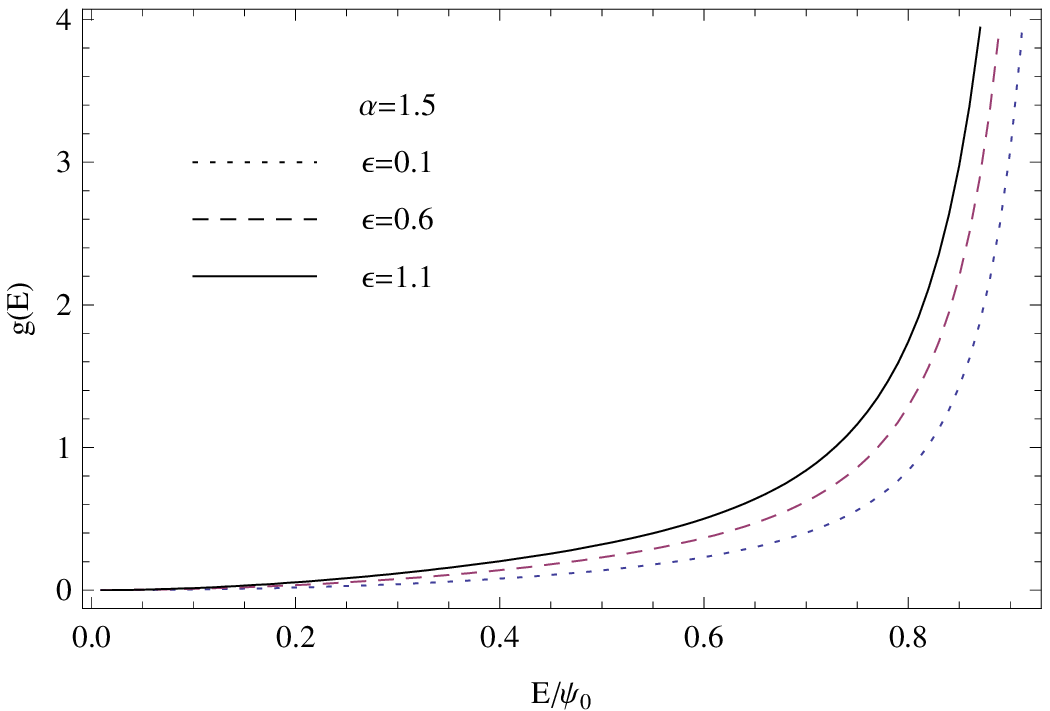}
\includegraphics[width=0.48\hsize]{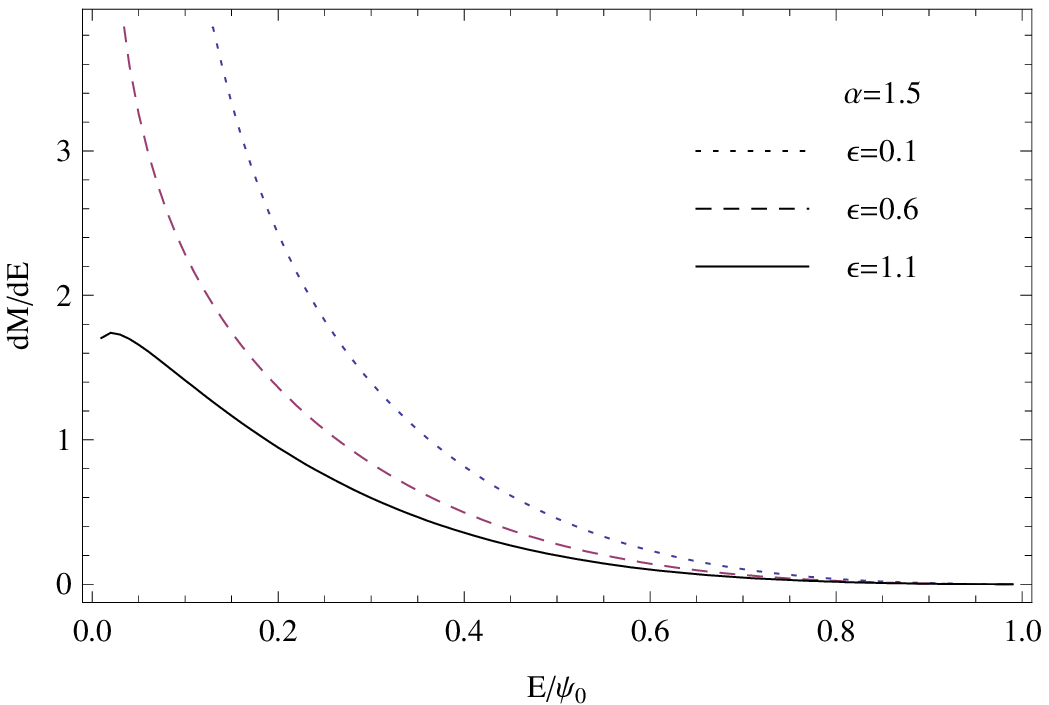}
\vspace*{8pt}
 \caption{\label{fig:dm}\footnotesize The energy part
of the distribution function (left panel) and the differential
energy distribution (right panel) of the New Generalized NFW model
with the constant anisotropy parameter $\beta=1/2$ and $\alpha=1.5$:
dotted lines ($\epsilon=0.1$, where the asymptotic behavior of
density profile is similar to the Moore model), dashed lines
($\epsilon=0.6$), solid lines ($\epsilon=1.1$). \label{f2}}
\end{figure*}

Now we can get $g(r_E)$ from Eq.~(\ref{eq:def}) and
Eq.~(\ref{eq:psi}):
\begin{gather}
\begin{split}&
g(r_E)=\left.\frac{B_1(r)} {4\pi\alpha\Gamma(-\alpha)[B_2(r) A_4(r)
+ B_3(r) A_5(r)]}\right|_{r=r_E}
\\&
\\&
B_1(r)=-(-2+\alpha-\epsilon)\ (-1+\alpha-\epsilon)\ [-1 + \alpha +
(2 + \epsilon)r ]\ r\ (1+r)^{-2+\alpha-\epsilon}
\\&
B_2(r)=(2-\alpha)+(4+2\epsilon-3\alpha)r+[(2+\epsilon)(1+\epsilon)-\alpha(3+2\epsilon)]r^2
\\&
B_3(r)=\alpha-2+(2\alpha\ - 3 -\epsilon) r
\\&
 A_4(r)=\ _2\tilde{F}_1[1 -\alpha,\ 1 -\alpha+\epsilon,\ 2 -\alpha,\ -r]
\\&
A_5(r)=\ _2\tilde{F}_1[-\alpha,\ 1-\alpha+\epsilon,\ 2 -\alpha,\ -r]
\end{split}
\end{gather}

Note that $B_1(r)=0$ if $\alpha-\epsilon=1, 2$. However, $g(r_E)$
doesn't approach zero when $\alpha-\epsilon\rightarrow 1, 2$. We
also find that $\alpha$ should be larger than $1$ as the
distribution function should be positive. We couldn't get the $r_E$
from $\psi(r_E)=E$ analytically but can get the numerical solutions.
$F(E,L)$ and $dM/dE$ are expressed as:
\begin{equation}
F(E,L)=\left.\frac{B_1(r)} {8\pi^3 L \alpha\Gamma(-\alpha)[B_2(r)
A_4(r) + B_3(r) A_5(r)]}\right|_{r=r_E}
\end{equation}

\begin{equation}
\frac{dM}{dE}=\left.\frac{r^2\ B_1(r)}
{2\alpha\Gamma(-\alpha)[B_2(r) A_4(r) + B_3(r)
A_5(r)]}\right|_{r=r_E}\ .
\end{equation}
Fig.~\ref{f1} and Fig.~\ref{f2} show $g(E)-E/\psi_0$ and
$dM/dE-E/\psi_0$ for different values of parameters $\alpha$ and
$\epsilon$. Here, $\psi_0=\psi(r=0)$.

\subsection{Generalized Moore profiles}
The profile of the Moore model\cite{M99} is shown as
$\rho\propto1/((r/r_s)^{1.5}(1+(r/r_s)^{1.5}))$. We consider one
generalized density profile of the Moore model:
\begin{equation}
\rho=C\frac{1}{(r/r_s)^\alpha(1+(r/r_s)^{3+\epsilon-\alpha})}.
\end{equation}

As in section 3.1, we still set $r_s=1$, $M_{tot}=1$ and $G=1$ here.
We can not express $C$ as a function of $\alpha$ and $\epsilon$
analytically but can calculate $C$ in a numerical way while
considering the condition $M_{tot}=\int_0^{\infty}4\pi
r^2\rho(r)dr=1$.

The relative potential is
\begin{gather}
\begin{split}&
\psi(r)=
 C[c_0 - \frac{4\pi r^{2-\alpha}
   \Gamma_1 \Gamma_2 A_6(r)}{(\alpha-2)(\alpha-3)} ]
\\&
\\&
\Gamma_1=\Gamma[\frac{5 - 2 \alpha + \epsilon}{3 - \alpha +
\epsilon}]
\\&
\Gamma_2=\Gamma[\frac{6 - 2 \alpha + \epsilon}{3 - \alpha +
\epsilon}]
\\&
A_6(r)=
\\&
\ _3\tilde{F}_2[\frac{-3+\alpha}{-3+\alpha-\epsilon},1,
\frac{-2+\alpha}{-3+\alpha-\epsilon};\frac{6-2\alpha+\epsilon}{3-\alpha+\epsilon},
\frac{5-2\alpha+\epsilon}{3-\alpha+\epsilon}; -r^{3 - \alpha +
\epsilon}]
 \label{eq:psi2}
\end{split}
\end{gather}
where $c_0$ is determined by
$\left.\psi(r)\right|_{r\rightarrow\infty}=0$ numerically. $\psi(r)$
cannot satisfy the condition
$\left.\psi(r)\right|_{r\rightarrow\infty}=0$ for some values of
$\alpha\geq2$. Therefore, just as in section 3.1, we only focus on
the situation of $\alpha=1 \sim 1.5$. Then $g(r_E)$ can be
calculated from Eq.~(\ref{eq:def}) and Eq.~(\ref{eq:psi2}):

\begin{gather}
\begin{split}&
g(r_E)=
 \left.-\frac{B_4(r)}{4\pi B_5(r) \Gamma_1}\ \frac{B_6(r)}
     {(-3 + \alpha)\ A_7(r)+ \Gamma_2 \ A_6(r)}\right|_{r=r_E}
\\&
\\&
A_7(r)=\ _2\tilde{F}_1[1, \frac{-2 + \alpha}{-3 +\alpha -\epsilon},
\frac{5 -2\alpha +\epsilon}{3 -\alpha + \epsilon}, -r^{3 -
\alpha+\epsilon}]
\\&
B_4(r)=\ (-1 + \alpha) r^\alpha + (2 + \epsilon) r^{3 + \epsilon}
\\&
B_5(r)=\ (r^\alpha + r^{3 + \epsilon})^2
\\&
B_6(r)=\ (6 -5\alpha+ \alpha^2) r^{-1 +\alpha}
\end{split}
\end{gather}
\begin{figure*}[h]
\includegraphics[width=0.482\hsize]{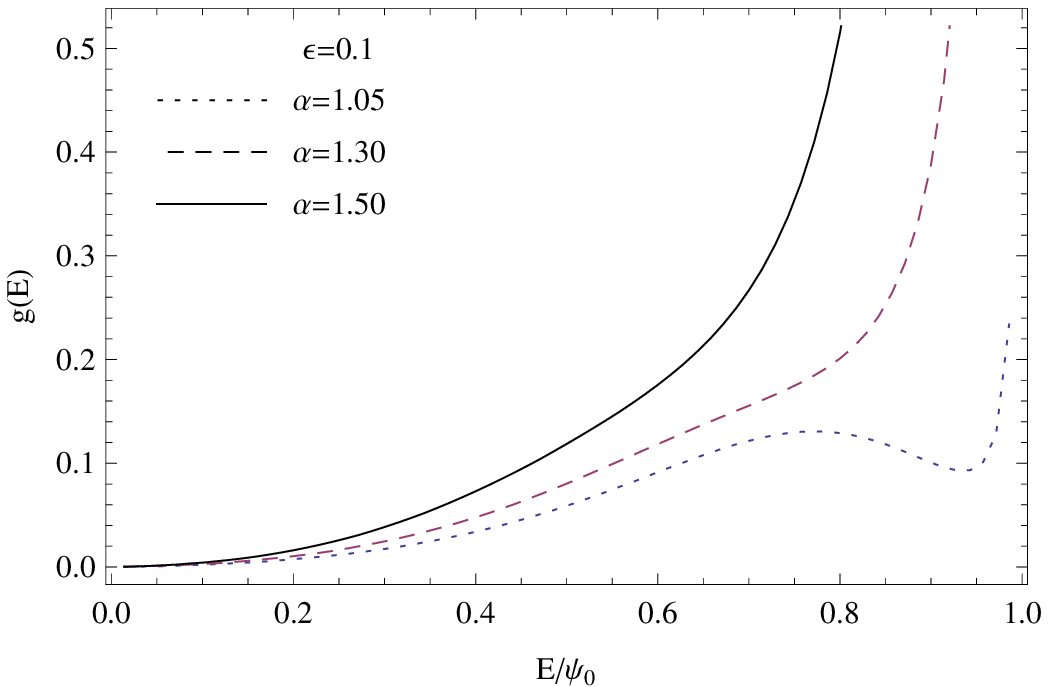}
\includegraphics[width=0.482\hsize]{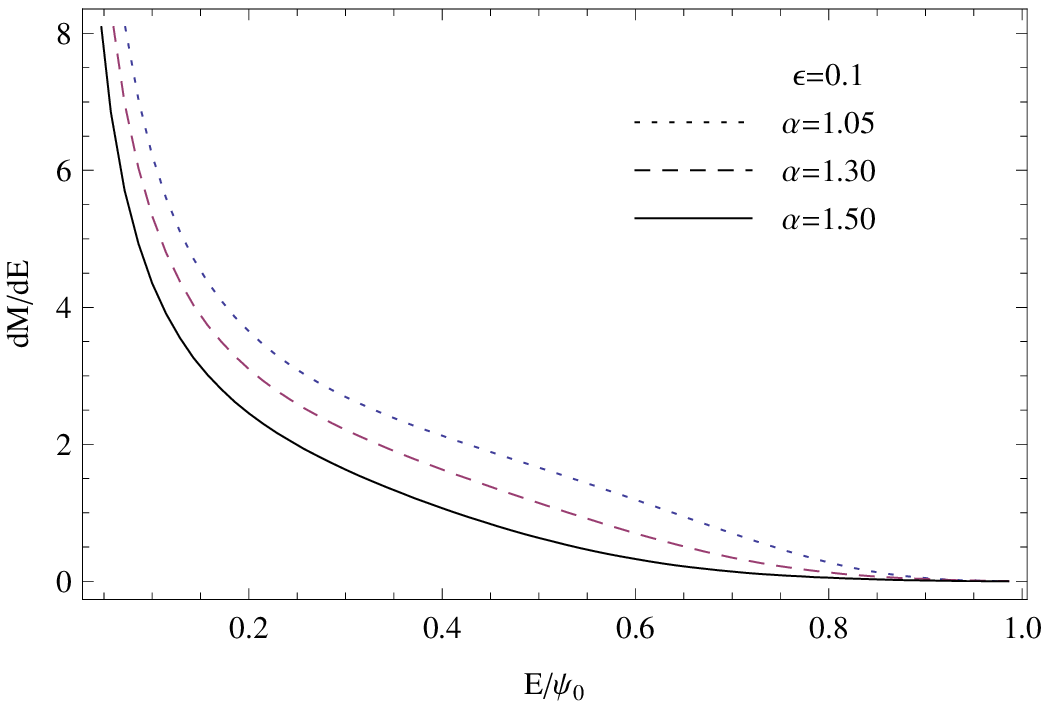}
\vspace*{8pt}
 \caption{\label{fig:gdf}\footnotesize The energy part
of the distribution function (left panel) and the differential
energy distribution (right panel) of the Generalized Moore model
with the constant anisotropy parameter of $\beta=1/2$ and
$\epsilon=0.1$: dotted lines ($\alpha=1.05$, where the asymptotic
behavior of density profile is similar to the NFW profile's, but DF
doesn't meet the condition $dg(E)/dE> 0$.), dashed lines
($\alpha=1.30$), solid lines ($\alpha=1.50$, closes to the Moore
model). \label{f3}}
\end{figure*}

\begin{figure*}[h]
\includegraphics[width=0.48\hsize]{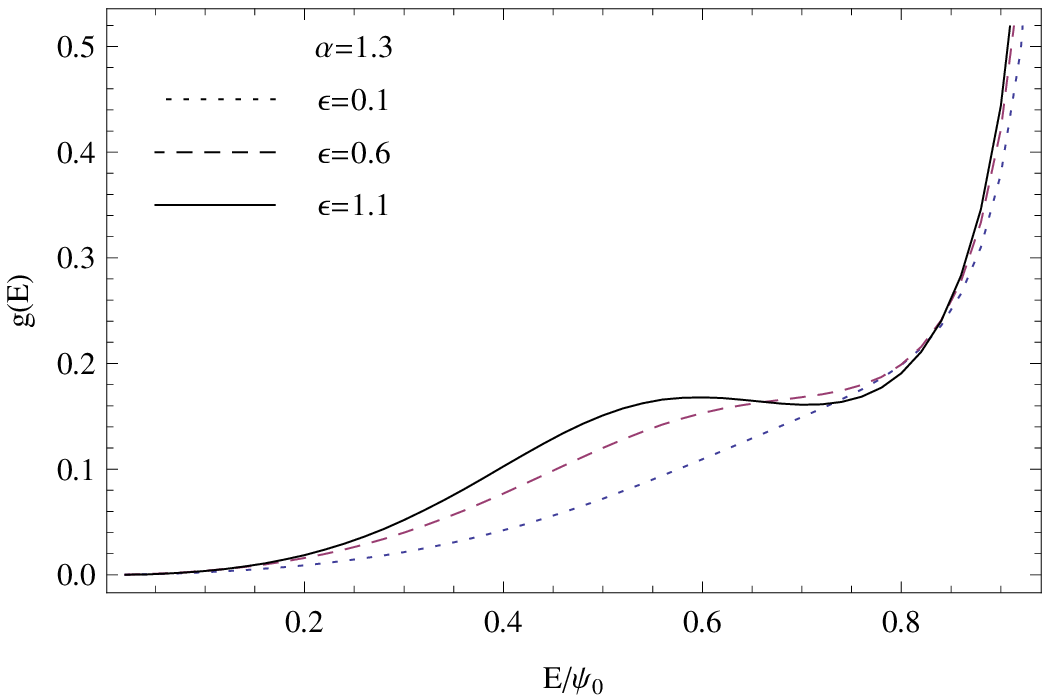}
\includegraphics[width=0.48\hsize]{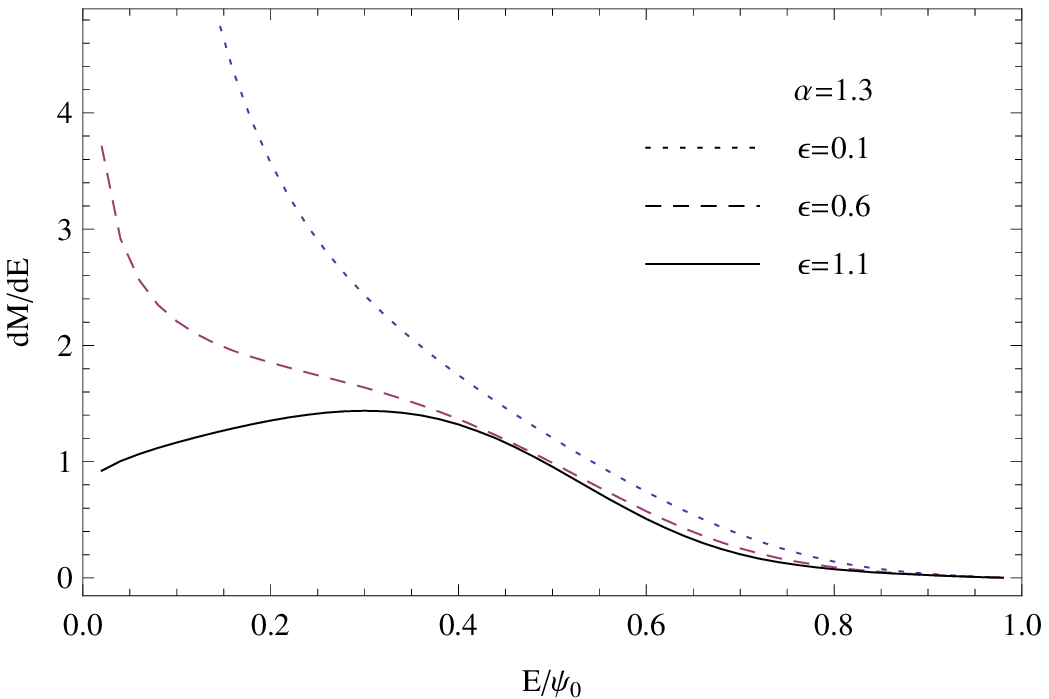}
\vspace*{8pt}
 \caption{\label{fig:gdf}\footnotesize The energy part
of the distribution function (left panel) and the differential
energy distribution (right panel) of the Generalized Moore model
with the constant anisotropy parameter of $\beta=1/2$ and
$\alpha=1.3$: dotted lines ($\epsilon=0.1$), dashed lines
($\epsilon=0.6$), solid lines ($\epsilon=1.1$, where DF doesn't meet
the condition $dg(E)/dE> 0$). \label{f4}}
\end{figure*}

In this kind of profile, we also calculate $r_E$ by numerical way.
The $F(E,L)$ and $dM/dE$ are expressed as:
\begin{gather}
\begin{split}&
F(E,L)=\
 \left.-\frac{B_4(r)}{8\pi^3 L\ B_5(r) \Gamma_1}\ \frac{B_6(r)}
     {(-3 + \alpha)\ A_7(r)+ \Gamma_2 \ A_6(r)}\right|_{r=r_E}
\\&
\frac{dM}{dE}=\ \left.-\frac{r^2\ B_4(r)}{2 B_5(r) \Gamma_1}\
\frac{B_6(r)}
     {(-3 + \alpha)\ A_7(r)+ \Gamma_2 \ A_6(r)}\right|_{r=r_E}\ .
\end{split}
\end{gather}
Fig.~\ref{f3} and Fig.~\ref{f4} show $g(E)-E/\psi_0$ and
$dM/dE-E/\psi_0$ for different values of parameters $\alpha$ and
$\epsilon$.

\begin{table}
  \tbl{\label{table1} Parameters $\epsilon$ and $\alpha_c$}
  {\begin{tabular}{@{}ccccccc@{}} \toprule
  $\epsilon=$ & \hspace{0.4cm} $0.01$ & \hspace{0.4cm}
  $0.1$ & \hspace{0.4cm} $0.2$
  & \hspace{0.4cm} $0.3$ & \hspace{0.4cm} $0.4$ & \hspace{0.4cm} $0.5$  \\
  $\alpha_c\simeq$ & \hspace{0.4cm} $1.110$ & \hspace{0.4cm}
  $1.127$   & \hspace{0.4cm} $1.146$ & \hspace{0.4cm} $1.166$ & \hspace{0.4cm} $1.186$ & \hspace{0.4cm} $1.209$
  \\ \colrule    $\epsilon=$ & \hspace{0.4cm} $0.6$ &
  \hspace{0.4cm}   $0.7$ & \hspace{0.4cm} $0.8$
  & \hspace{0.4cm} $0.9$ & \hspace{0.4cm} $1.0$ \\   $\alpha_c\simeq$ & \hspace{0.4cm} $1.230$ &   \hspace{0.4cm} $1.253$
  & \hspace{0.4cm} $1.276$ & \hspace{0.4cm}
  $1.299$ & \hspace{0.4cm} $1.323$ \\ \botrule
 \end{tabular}\label{ta1}}
 \end{table}

\begin{figure}[h]
\centerline{\psfig{file=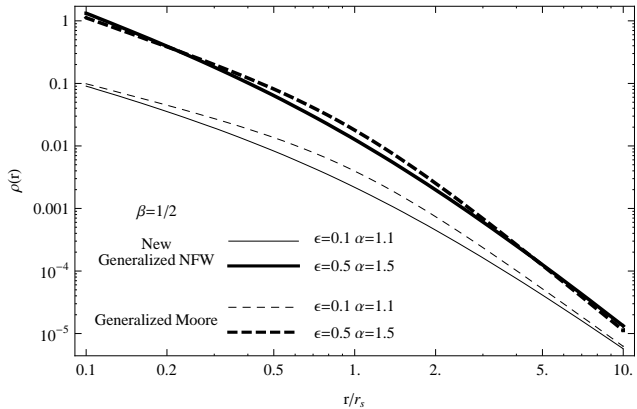,width=8.5cm}} \vspace*{8pt}
\caption{We set $G=1, M_{tot}=1$ and $\beta=1/2$. Two solid lines:
New Generalized NFW profiles with different parameters. Two dashed
lines: Generalized Moore profiles with different parameters.
\label{f5}}
\end{figure}

In this model, a basic stability condition\cite{Ant62,DFB71,BT87}
for the spherical stellar system $dF/dE > 0$ is equivalent to the
condition $dg(E)/dE> 0$ (note: the "$E$" here is not the energy of a
particle but the binding energy). We note that spherical systems
with certain $\alpha$ and $\epsilon$ may not meet this condition.
This condition requires that the parameter $\alpha_i$ should be
larger than a critical value $\alpha_c$ for the given parameter
$\epsilon_i$. We calculate $\alpha_c$ numerically and show them in
Table~\ref{ta1}. The difference of the two generalized profiles is
shown in Fig.~\ref{f5}. One can see that the parameter $\alpha$
controls the asymptotic density profile in the inner parts, while
$\epsilon$ is responsible for the asymptotic behavior of the outer
profile.

\section{Discussion and Conclusions}

N-body simulations show that the anisotropy parameter
$\beta\approx0.5$ in the outer part of dark matter halos. The
distribution functions with constant anisotropy parameter
$\beta=1/2$ and the corresponding differential energy distributions
can be calculated from the formulae~(\ref{eq:def}). N-body
simulations also show that the logarithmic slope of density profile
$\gamma=1\sim1.5$ in the inner part of the dark matter halo
($\rho\sim r^{-\gamma}$).

Many authors work at construction of the dark matter halo models
which are better to be analytical, simple, realistic and having
flexible anisotropy profile. Unfortunately, it is difficult to meet
these requirements at the same time.

In this paper, we only consider the models with constant anisotropy
parameter $\beta=1/2$ and then calculate the DFs and DEDs of the two
models, the new generalized NFW model and the generalized Moore
model. Both have two parameters $\epsilon$ and $\alpha$ for
determination of the asymptotic behavior of density profile in the
inner and outer parts of the halo. With these two parameters, our
models can cover many relatively realistic density profiles which
come from N-body simulations. A physical postulate requires that DFs
should be positive in the phase space, and our DFs satisfy this
basic condition. A basic stability condition for a spherical stellar
system $dF/dE>0$ is the sufficient condition for the isotropic case,
but is no more a sufficient condition for the anisotropic
one.\cite{Ant62,DFB71,BT87} However, we still can use this condition
to do some rough discussion. In the new generalized NFW model, all
DFs with $\epsilon>0$ and $\alpha=1\sim1.5$ meet the condition
$dF/dE>0$. In the generalized Moore model, we find that not all DFs
with $\epsilon>0$ and $\alpha=1\sim1.5$ satisfy this condition. In
order to satisfy the condition $dF/dE>0$, $\alpha_i$ should be
larger than a critical value $\alpha_c$ for a given parameter
$\epsilon_i$.

We have not dealt with the stability of our models in details in our
paper. Although it is difficult to determine the stability domain in
the parameter space of the halo model, the study of stability of the
stellar system and dark matter halo is very important and some works
have already been done \cite{De88,Meza97,Meza02,BVDD07}. To be more
realistic, a more complicated case, the axisymmetric model should be
considered\cite{HQ93,HB01,FVD02,JO07}. Construction of the
polycomponent models is also necessary as the stellar component or
the central black hole\cite{MF01,BD04,BDB05} usually combines with
the dark matter component and the two-component models with the
stellar and dark matter components were constructed by
Ciotti\cite{Ci96}.

Although the density profiles of our models are realistic enough and
we assume that the anisotropy parameter $\beta=1/2$ from the
viewpoint of the simulation, it is still not realistic enough for
the anisotropy profile. An $\&$ Evans\cite{AE06} explored the model
whose DF has the form as below,
\begin{equation}
F(E,L)=\sum_i L^{-2\beta_i}f_i(E).
\end{equation}
DF is superposition of two or more terms here and this case is more
complicated, nevertheless, this kind of DF has more flexible
anisotropy profile and it is revealing for our further work to
pursue a greater variety of anisotropic behavior of the dark matter
halo models.

\section*{Acknowledgments}
DM thanks Prof. M. Baes, Drs. L. M. Cao and H. Li for useful
discussions and kind help. We are grateful for an anonymous referee
for his/her helpful and constructive comments to improve the
manuscript. This work is supported by the Scientific Research
Foundation for the Returned Overseas Chinese Scholars, State
Education Ministry of China, and by the Chinese Academy of Sciences
under Grant No. KJCX3-SYW-N2.

\appendix

\section{Hypergeometric Function\protect\footnote{See
http://mathworld.wolfram.com/}}

$_p\tilde{F}_q$ is the regularized hypergeometric function which is
defined as
\begin{equation}
\ _p\tilde{F}_q(a_1,...,a_p;b_1,...,b_q;z)\equiv\frac{\
_pF_q(a_1,...,a_p;b_1,...,b_q;z)}{\Gamma(b_1)...\Gamma(b_q)}\ ,
\end{equation}
where $\Gamma(z)$ is a gamma function. And $
_pF_q(a_1,...,a_p;b_1,...,b_q;z)$ is the generalized hypergeometric
function:
\begin{equation}
\ _pF_q(a_1,...,a_p;b_1,...,b_q;z)=\sum^{\infty}_{k=0}\frac{(a_1)_k\
(a_2)_k\ ...\ (a_p)_k}{(b_1)_k\ (b_2)_k\ ...\ (b_q)_k}\
\frac{z^k}{k!}\ ,
\end{equation}
where $(a)_k$ is the Pochhammer symbol,
\begin{equation}
(a)_k\equiv\frac{\Gamma(a+k)}{\Gamma(a)}=a(a+1)...(a+k-1)\ .
\end{equation}
The specific hypergeometric functions can be calculated by
mathematical software packages such as Mathematica.



\begin{thebibliography}{0}    

\bibitem{NFW95} J. F. Navarro, C. S. Frenk and S. D. M. White, {\it Mon. Not. R.
Astron. Soc.} {\bf 275} (1995) 720.

\bibitem{NFW96} J. F. Navarro, C. S. Frenk and S. D. M. White, {\it Astrophys. J.}
{\bf 462} (1996) 563.

\bibitem{M99}
B. Moore, T. Quinn, F. Governato, J. Stadel and G. Lake {\it Mon.
Not. R. Astron. Soc.} {\bf 310} (1999) 1147.

\bibitem{J2000}
Y. P. Jing and Y. Suto, {\it Astrophys. J.} {\bf 529} (2000) L69.

\bibitem{AR99}
V. Avila-Reese, C. Firmani, A. Klypin and A. V. Kravtsov {\it Mon.
Not. R. Astron. Soc.} {\bf 310} (1999) 527.

\bibitem{HM06}
S. H. Hansen and B. Moore, {\it New Astron.} {\bf 11} (2006) 333.

\bibitem{HS06}
S. H. Hansen and J. Stadel, {\it J. Cosmol. Astropart. P.} {\bf 05}
(2006) 014.

\bibitem{Bi80}
J. J. Binney, {\it Mon. Not. R. Astron. Soc.} {\bf 190} (1980) 873.

\bibitem{Ed16}
A. S. Eddington, {\it Mon. Not. R. Astron. Soc.} {\bf 76} (1916)
572.

\bibitem{M63}
R. W. Michie, {\it Mon. Not. R. Astron. Soc.} {\bf 125} (1963) 127.

\bibitem{King66}
I. R. King, {\it Astron. J.} {\bf 71} (1966) 64.

\bibitem{De86}
H. Dejonghe, {\it Phys. Rep.} {\bf 133} (1986) Nos 3 - 4.

\bibitem{De87}
H. Dejonghe, {\it Mon. Not. R. Astron. Soc.} {\bf 224} (1987) 13.

\bibitem{EA06}
N. W. Evans and J. H. An, {\it Phys. Rev. D} {\bf 73} (2006) 023524.

\bibitem{Os79}
L. P. Osipkov, {\it Pis'ma Astron. Zh.} {\bf 5} (1979) 77.

\bibitem{Os79tr}
L. P. Osipkov, {\it Soviet Astron. Lett.} {\bf 5} (1979) 42.

\bibitem{Me85}
D. Merritt, {\it Astron. J.} {\bf 90} (1985) 1027.

\bibitem{Cu91}
P. Cuddeford, {\it Mon. Not. R. Astron. Soc.} {\bf 253} (1991) 414.

\bibitem{CP92}
L. Ciotti and S. Pellegrini, {\it Mon. Not. R. Astron. Soc.} {\bf
255} (1992) 561.

\bibitem{BV07}
M. Baes and E. van Hese, {\it Astron. Astrophys.} {\bf 471} (2007)
419.

\bibitem{TLS06}
C. Tonini, A. Lapi, and P. Salucci, {\it Astrophys. J.} {\bf 649}
(2006) 591.

\bibitem{Pl11}
H. C. Plummer, {\it Mon. Not. R. Astron. Soc.} {\bf 71} (1911) 460.

\bibitem{Ve79}
U. I. K. Veltmann, {\it Astron. Zh.} {\bf 56} (1979) 976.

\bibitem{He90}
L. Hernquist, {\it Astrophys. J.} {\bf 356} (1990) 359.

\bibitem{BD02}
M. Baes and H. Dejonghe, {\it Astron. Astrophys.} {\bf 393} (2002)
485.

\bibitem{Deh93}
W. Dehnen, {\it Mon. Not. R. Astron. Soc.} {\bf 265} (1993) 250.

\bibitem{Tre94}
S. Tremaine, D. O. Richstone, Y. Byun, A. Dressler, S. M. Faber, C.
Grillmair, J. Kormendy and T. R. Lauer, {\it Astron. J.} {\bf 107}
(1994) 634.

\bibitem{BDB05}
M. Baes, H. Dejonghe and P. Buyle, {\it Astron. Astrophys.} {\bf
432} (2005) 411.

\bibitem{BHD07}
P. Buyle, C. Hunter and H. Dejonghe, {\it Mon. Not. R. Astron. Soc.}
{\bf 375} (2007) 773.

\bibitem{Too82}
A. Toomre, {\it Astrophys. J.} {\bf 259} (1982) 535.

\bibitem{AE05}
J. H. An and N. W. Evans, {\it Astron. Astrophys.} {\bf 444} (2005)
45.

\bibitem{Wi00}
L. M. Widrow, {\it Astrophys. J. Suppl. Ser.} {\bf 131} (2000) 39.

\bibitem{AE06}
J. H. An and N. W. Evans, {\it Astron. J.} {\bf 131} (2006) 782.

\bibitem{WE99}
M. I. Wilkinson and N. W. Evans, {\it Mon. Not. R. Astron. Soc.}
{\bf 310} (1999) 645.

\bibitem{Ant62}
V. A. Antonov, {\it Vestn. Leningr. Univ.} {\bf 7} (1962) 135.

\bibitem{DFB71}
J. P. Doremus, M. R. Feix and G. Baumann, {\it Phys. Rev.Lett.} {\bf
26} (1971) 725.

\bibitem{BT87}
J. Binney and S. Tremaine, {\it Galactic Dynamics}, (Princeton
University Press, Princeton, 1987).

\bibitem{De88}
H. Dejonghe and D. Merritt, {\it Astrophys. J.} {\bf 328} (1988) 93.

\bibitem{Meza97}
A. Meza and N. Zamorano, {\it Astrophys. J.} {\bf 490} (1997) 136.

\bibitem{Meza02}
A. Meza, {\it Astron. Astrophys.} {\bf 395} (2002) 25.

\bibitem{BVDD07}
P. Buyle, E. Van Hese, S. De Rijcke and H. Dejonghe, {\it Mon. Not.
R. Astron. Soc.} {\bf 375} (2007) 1157.

\bibitem{HQ93}
C. Hunter and E. Qian, {\it Mon. Not. R. Astron. Soc.} {\bf 262}
(1993) 401.

\bibitem{HB01}
K. Holley-Bockelmann, J. C. Mihos, S. Sigurdsson and L. Hernquist,
{\it Astrophys. J.} {\bf 549} (2001) 862.

\bibitem{FVD02}
B. Famaey, K. Van Caelenberg and H. Dejonghe, {\it Mon. Not. R.
Astron. Soc.} {\bf 335} (2002) 201.

\bibitem{JO07}
Z. L. Jiang and L. Ossipkov, {\it Mon. Not. R. Astron. Soc.} {\bf
379} (2007) 1133.

\bibitem{MF01}
D. Merritt and L. Ferrarese, {\it Mon. Not. R. Astron. Soc.} {\bf
320} (2001) 30.

\bibitem{BD04}
M. Baes and H. Dejonghe, {\it Mon. Not. R. Astron. Soc.} {\bf 351}
(2004) 18.

\bibitem{Ci96}
L. Ciotti, {\it Astrophys. J.} {\bf 471} (1996) 68.

\end{thebibliography}
\end{document}